\magnification=\magstep1
\hsize=159.2 true mm
\vsize=236 true mm
\baselineskip=13pt
\parskip = 4pt
\overfullrule=0pt
\def\sec#1{\vbox{\medskip\noindent{\bf#1}\smallskip}\nobreak}
\headline={\ifnum\pageno=1\leftline{\sl
An invited talk presented at StatPhys-Taipei97, 4--11 August 1997}\else\hfill\fi}
\topglue 2cm
\font\bbf=cmbx12

\centerline{\bbf Random Sequential Adsorption,} 
\centerline{\bbf Series Expansion and Monte Carlo Simulation}
\bigskip
\centerline{\bf Jian-Sheng Wang}
\smallskip
\centerline{\sl Department of Computational Science,}
\centerline{\sl The National University of Singapore, Singapore, 119260}

\bigskip
\bigskip
\centerline{Abstract}

{ \narrower 
Random sequential adsorption is an irreversible surface deposition of
extended objects.  In systems with continuous degrees of freedom
coverage follows a power law, $\theta(t) \approx
\theta_J - c\, t^{-\alpha}$, where the exponent $\alpha$ depends
on the geometric shape (symmetry) of the objects.  Lattice models give 
typically exponential saturation to jamming coverage.  We discuss how such 
function $\theta(t)$ can be computed by series expansions  and analyzed with 
Pad\'e approximations.  We consider the applications of 
efficient Monte Carlo computer simulation method (event-driven method)
to random sequential adsorptions with high precision and at very 
long-time scale. 



}

\bigskip
\bigskip

\sec{1.~Introduction}

Random sequential adsorption$^{1,2}$ (RSA) refers to a type of
irreversible adsorptions of spatially extended objects.  Two features
characterize RSA: (1) the objects, once on the surface, are
permanently stick to the surface without further thermal relaxation;
(2) only a single layer is deposited, and the objects have hardcore
volume exclusion.

A standard example$^{3}$ of RSA is the adsorption of spherical
macromolecular particles in solution on a glass surface.  Some
surface reaction dynamics may be described by RSA on lattices.  In
comparison with a corresponding equilibrium system of particles, RSA
shows unusual properties, most notably the existence of a jamming
coverage and a super-exponential correlation of particle density.

One of the central theoretical problems in RSA is the computation of
the coverage $\theta(t)$.  Such problems are intractable for exact
solutions except perhaps some one-dimensional systems.$^{4,5}$ Series
expansion and Monte Carlo simulation offer semi-numerical or numerical
results which can be improved systematically.  They are perhaps the best
methods available.  In this article, we describe some of the advanced
techniques of attaining that goal.

\sec{2.~Series Expansions}

Earlier calculations$^{2,6}$ in RSA were based on truncation of an
infinite hierarchy of rate equations together with mean-field
type approximations (and shielding property) to close the equations.
It is only recently that these equations are systematically used,
resulting very long series.$^{7-10}$

To illustrate the series expansion method,$^9$ let us consider the
deposition of dimer on square lattice.  The basic quantity of
interest in computing a series is the probability $P(A)$ that the
given set $A$ of sites are empty; we do not care other sites being
occupied or empty.  It is a marginal probability of the full
probability distribution.  Such probability can only decrease.  The
rate of decrease is proportional to the probability that the current
configuration $A$ can be destroyed by depositing a dimer with at least
one site in $A$.  Thus we have
$$ - { d P(A) \over dt } \propto\mskip-24mu
\sum_{{\rm ways\ of\ destroying\ } A}\mskip-30mu  P(A'). \eqno(1)$$
where the summation is over all possible ways of depositing a dimer at a
pair of empty sites such that at least one site is in $A$; $A'=A$ if
two of the dimer sites of a deposition attempt are within $A$, or $A'$
is one site more than $A$ so that the deposition with one site in $A$ and
one site outside $A$ can be carried out.  The proportionality constant
sets the time scale.  Without loss of generality, we can take it to be
unity.  The first two equations look like these:
$$ - { dP({\rm o})\over dt }  = 4\,P({\rm oo}),  \eqno(2a) $$
$$ - { dP({\rm oo})\over dt } = P({\rm oo}) + 2\,P({\rm ooo}) 
+ 4\,P(\vbox{\hbox{\lower 7pt \hbox{o}}\hbox{oo}}). \eqno(2b) $$
There are four ways to destroy a single empty site, provided that the
nearest neighbor site is also empty.  Using the assumption that
initial conditions are lattice symmetry invariant, we can write them simply
as $4 P({\rm oo})$.  The second equation is derived similarly.

For a general discussion, we write the rate equations symbolically as 
$$ { dP(A)\over dt}  = {\cal L} P(A), \eqno(3)$$
where $\cal L$ is a linear operator defined by
$$ {\cal L} P(A) = \sum_{A'} c_{A'} P(A'). \eqno(4) $$
The $n$-th derivative is then 
$$ { d^n P(A)\over dt^n } 
= {\cal L}^n P(A), \qquad {\rm with\ }\quad P(A')\big|_{t=0} = 1 
\quad {\rm\ for\ all\ }A'. \eqno(5)$$
The initial conditions are such that all sites are empty at $t=0$.

The above scheme can be implemented on computer as follows:
starting from some initial pattern (or called configuration) $A^0$, 
we apply ${\cal L}$ to get
$A^1_j$, from which we obtain the first derivative at $t=0$, as $
dP(A^0)/dt|_{t=0} = \sum_j c^1_j P(A^1_j)|_{t=0} = \sum_{j} c^1_j.$
The second derivative is obtained recursively by applying ${\cal L}$
again to the new patterns as
$$ { d^2 P(A^0)\over dt^2}  = \sum c^1_j{ dP(A^1_j)\over dt}, \eqno(6a)$$
$$ { dP(A^1_j)\over dt} = {\cal L} P(A^1_j) = \sum_{k} c^2_k  P(A^2_k).\eqno(6b)$$
The superscript 2 in $c_k^2$ refers to the second generation of patterns.
The relationship of the patterns and the rate equations can be
represented as a tree with arbitrary number of branches.  Each node of
a tree contains a pattern $A$, and its associated derivatives $d^i
P(A)/dt^i$ for $i = 0, 1,2, \ldots, n$.  Each node, if its derivative
has been computed, has a list of pointers to the children nodes and
associated coefficients $c_j$.  Thus a node together with its children
symbolically represents one rate equation.

If the tree nodes are traversed in a  depth-first manner, then we need not store all
the nodes at the same time.  The memory requirement is proportional to
the number of levels of the tree.  Simple counting method usually uses
this strategy.  The same node may have to be expanded many
times, and repeated work has to be done.  However, minimum book-keeping is
needed.

Computation can be performed faster if we do not repeat the same expansion and
calculation.  However, it is necessary then that all nodes are kept;
the tree can be generated in a breadth-first fashion one level at a
time.  Equivalent nodes due to lattice symmetry are treated as the
same node.  If a particular node is already generated before, a
pointer reference is made to the existing node.  Such strategy is
known as dynamic programming in computer science.

The most practical and efficient algorithm is a combination of several
methods.  To conserve memory, depth-first expansion after every $D$
levels is made; while the highest levels use a simple counting
algorithm which uses a negligible amount of memory.  

The above method is applicable to a wide range of problems, such as
diffusion-reaction models, Ising relaxation dynamics, and other
dynamic processes described by a hierarchy of rate equations.

\noindent Summary of Results

Series$^{10}$ for the dimer on square lattice (17 orders), monomer with
nearest neighbor exclusion on square lattice (20 orders), 
dimer on honeycomb lattice (20
orders), and monomer with nearest neighbor exclusion on 
honeycomb lattice (23 orders) are
obtained.  To our knowledge, there series are the longest known.

The discrete counting problem for RSA on lattice can be generalized to
RSA on continuum, where summations are replaced by integrals.  We note
that these integrals can be related to the cluster integrals with a
diagrammatic rule similar to the Mayer theory for fluid.  A five-order
series for disks and a seven-order series for aligned squares or cubes 
are available.$^{7,11}$

\noindent Pad\'e Analysis of Series

Given a finite number of terms in a series, how can we say anything about
the whole time domain $t$ from 0 to $\infty$?  Clearly it is not
possible in general.  Our experience with RSA series suggests that we
can give reliable estimates from the series for all times.$^7$

The method of Pad\'e approximants$^{12,13}$ is very useful in this
respect.  Given a series $f(x)$ to order $L$, we determine two
polynomials $P_N(x)$ and $Q_D(x)$ of degree $N$ and $D$ respectively,
such that
$$ f(x) - { P_N(x)\over Q_D(x) } = O(x^{L+1}), \qquad N+D\leq L. \eqno(7)$$
It is a powerful way of extending the domain of convergence of the
original series.  To accelerate the convergence further, new variables
are introduced,$^7$ e.g.,
$$ s = 1 - \exp\bigl(-b ( 1 - e^{-t})\bigr). \eqno(8)$$
The functional form is chosen in such a way so that the series
in the new variable $s$ is most closely resemble the asymptotic
behavior of the coverage $\theta(t)$ at large $t$.  The above specific
form is encouraged by the exact solution of the one-dimensional dimer
problem.  In fact we have $\theta = s$ with $b=2$ in such case.  The
convergence among various Pad\'e approximants is improved greatly by the 
transformation.

Here is an example of Pad\'e approximant for the dimer on square lattice
$$\eqalignno{ & \Big(2.962963\,s + 0.03206897\,{s^2} - 2.195246\,{s^3} 
 - 1.073721\,{s^4}  +  0.9207869\,{s^5} + 0.5556586\,{s^6} & \cr
 & - 0.04386743\,{s^7} - 0.05303456\,{s^8} \Big)\Big/\Big( 
   1 + 1.733045\,s - 0.2568919\,{s^2} - 1.942572\,{s^3} &  \cr
 &   - 0.5852424\,{s^4} + 0.7908992\,{s^5} + 0.4421557\,{s^6} - 
     0.0493306\,{s^7} - 0.0513337\,{s^8} \Big), & (9)\cr}$$
where $s$ is given by Eq.~(8) with $b = 1.35$.  The result is stable
against variation in $b$.  Error can be estimated from the convergence
of various Pad\'e approximants and Monte Carlo simulation results.
The above Pad\'e approximant is accurate to
$10^{-5}$ for all $t\geq 0$.  Such a high degree of accuracy embedded
in a simple formula is remarkable.

\sec{3.~Monte Carlo Simulation}

Monte Carlo simulation$^{14-16}$ is a simple and useful method to get
a quick result on an otherwise difficult problem by analytical means.  RSA
model can be simulated rather easily with a simple program.   Our
emphasis here is how to make Monte Carlo simulation more
efficient.  During late stage of simulation, most of the sites are
already occupied or blocked and few adsorption attempts may succeed.
The dynamics thus becomes very slow, especially for deposition
on continuum.

To overcome such slow dynamics, event-driven algorithms$^{16-19,22}$ are
devised where the deposition attempts can be chosen in such a way so that it is
accepted with a high probability.

To illustrate the method, let us look at the dimer RSA on square lattice
again.  At late stage, the number of places where a dimer can be
placed is small.  We can afford to enumerate them and store them in a
list.  In the next stage of simulation, we pick a possible dimer
position from the list randomly, and make the deposition attempt.  The
difference between the new method and the original is that time step has to
be different.  In particular, each deposition attempt taken from
the list should advance the time step stochastically by
$$  \delta t = {1 \over N_{tot}} \left( 
\Bigl\lfloor { \ln \xi \over \ln (1-r) } \Bigr\rfloor
+ 1 \right), \eqno(10)$$
where $N_{tot}$ is total number of lattice sites, and $r$ is the ratio of
the number of ways of possible depositions on the list to the total
number of ways of depositions.  $\xi$ is a uniformly distributed 
random number between 0 and 1.  The original method has a constant 
time step $1/N_{tot}$ 

Similar idea is used for RSA of disks on continuum,$^{18}$ where the
area is divided into small squares.  Each square is classified as
available or unavailable for further deposition.  The classification
is nontrivial, but can be done efficiently in two dimensions.  The
new algorithm enable us to simulate the system until jamming state in 
finite computer time.
We confirmed the Feder's law$^{14}$
$$ \theta(t) - \theta_J \propto t^{-1/2}, \eqno(11)$$
and also obtained a very precise value for the jamming coverage $\theta_J
\approx 0.547069$.

An important application of the event-driven algorithms in
RSA is the deposition of more complex objects$^{19}$ like polymers and
macromolecules represented as random walks.  A lattice RSA model that
we have proposed describes an irreversible adsorption of polymer
chains in good solvent (self-avoiding random walks).  The coverage
is a very slow function of time.  Thus a straightforward simple
algorithm can only probe the short-time behavior.  The event-driven
algorithm requires to list all the available ways of deposit polymers
on the lattice.  Clearly such enumeration is not possible for long
chains due to the huge number of possible configurations a polymer can
take.  Thus we only enumerate the first few steps of the chain, and
sample remaining sites stochastically.  The length of the stored
segments of the chains is dictated by the computer memory available.
The event-driven method substantially extends the time scale, and is
able to simulate the process over 12 decades in time.  Our numerical results
suggest that the coverage as a function of time obeys
$$ \theta - \theta_J \propto t^{-2/N}, \eqno(12)$$
in a broad time domain for long chains, where $N$ is the length of the 
self-avoiding walks.  The jamming coverage is consistent with 
$\theta_J \propto N^{-0.1}$.

When the objects in RSA deposition are allowed to relax, specifically,
a random diffusion on the surface, the picture changes a lot. In
particular, a full coverage can be reached (depends on the geometry of
the objects).  RSA with diffusional relaxation is studied by Monte
Carlo simulations$^{20,21}$ as well as series expansions.$^{22}$
Unlike simple RSA, the late stage dynamics typically obeys a power law
even on discrete lattices.  Some of the peculiar behavior$^{21}$ have
not been understood very well.  Other research direction is the adsorption
involving multi-layers.$^{23,24}$ It is interesting to investigate new
phenomena and to develop efficient methods for such more complex
problems.

\sec{Acknowledgements}

The author would like to thank R.~Dickman, C.-K.~Gan, P.~Nielaba, and 
V.~Privman for many of the collaborative work.  This work is supported
in part by an Academic Research Grant, No.~PR950601, of National
University of Singapore.

\bigskip
\sec{References}

\newcount\itemno 
\itemno=0
\def\itm{\global\advance\itemno 
by1\item{\the\itemno.}} 

\frenchspacing
\parskip=2pt

\itm M. C. Bartelt and V. Privman, Int. J. Mod. Phys. B {\bf 5}, 2883 (1991).

\itm J. W. Evans, Rev. Mod. Phys., {\bf 65}, 1281 (1993).

\itm G. Y. Onoda and E. G. Liniger, Phys. Rev. A. {\bf 33}, 715 (1986).

\itm E. R. Cohen and H. Reiss,  J. Chem. Phys. {\bf 38}, 
680 (1963).

\itm J. J. Gonzalez, P. C. Hemmer, and J. S. H\o ye, Chem. Phys. {\bf 3},
228 (1974).

\itm R. S. Nord and J. W. Evans, J. Chem. Phys.
{\bf 82}, 2795 (1985).

\itm R. Dickman, J.--S. Wang, and I. Jensen, 
J. Chem. Phys. {\bf 94}, 8252 (1991).

\itm A. Baram and M. Fixman, J. Chem. Phys. {\bf 103}, 1929 (1995).

\itm C.--K. Gan and J.--S. Wang,
J. Phys. A, Math. Gen., {\bf 29}, L177 (1996).

\itm C.--K. Gan and J.--S. Wang, unpublished.

\itm B. Bonnier, M. Hontebeyrie, and C. Meyers, 
Physica A, {\bf 198}, 1 (1993).

\itm G. A. Baker, Quantitative Theory of Critical Phenomena,
Academic Press (1990).
 
\itm J. Adler, Annual Reviews of Computational Physics IV, pp.241-266
(1996).

\itm J. Feder, J. Theor. Biol. {\bf 87}, 237 (1980).

\itm E. L. Hinrichsen, J. Feder, and T. J\o ssang, J. Stat. Phys.
{\bf 44}, 793 (1986).

\itm B. J. Brosilow, R. M. Ziff, and R. D. Vigil, Phys. Rev. A {\bf 43},
631 (1991). 

\itm V. Privman, J.--S. Wang, and P. Nielaba,
Phys. Rev. B {\bf 43}, 3366 (1991).

\itm J.--S. Wang, Int. J. Mod. Phys. C {\bf 5}, 707 (1994).

\itm J.--S. Wang and R. B. Pandey, Phys. Rev. Lett.
{\bf 77}, 1773 (1996).

\itm J.--S. Wang, V. Privman, and P. Nielaba, 
Mod. Phys. Lett. B {\bf 7}, 189 (1993).

\itm J.--S. Wang, P. Nielaba, and V. Privman, 
Physica A, {\bf 199}, 527 (1993). 

\itm C.--K. Gan and J.--S. Wang, 
Phys. Rev. E. {\bf 55}, 107 (1997).

\itm P. Nielaba, V. Privman, and J.--S. Wang, 
J. Phys. A: Math. Gen. {\bf 23}, L1187 (1990).

\itm P. Nielaba, V. Privman, and J.--S. Wang, 
in {\it Computer Simulation Studies in Condensed-Matter Physics VI},
D. P. Landau, K. K. Mon, H.-B. Sch\"uttler, eds., Springer Proceedings
in Physics, Vol.~76, p.~143 (Springer-Verlag, Heidelberg, Berlin,
1993).

\bye